\documentclass[a4paper,11pt]{article}
\usepackage{jcappub}
\usepackage{natbib}
\usepackage{parskip}
\usepackage{hyperref}
\usepackage{epstopdf}
\usepackage{color}
\usepackage{gensymb}
\usepackage[usenames,dvipsnames]{xcolor}
\usepackage{epsfig}
\graphicspath{{images/}}

\newcommand{\bi}{\begin{itemize}}
\newcommand{\ei}{\end{itemize}}
\newcommand{\beq}{\begin{equation}}
\newcommand{\eeq}{\end{equation}}
\newcommand{\bea}{\begin{eqnarray}}
\newcommand{\eea}{\end{eqnarray}}
\newcommand{\bqu}{\begin{quote}}
\newcommand{\equ}{\end{quote}}
\newcommand{\bctr}{\begin{center}}
\newcommand{\ectr}{\end{center}}
\newcommand{\bd}{\begin{description}}
\newcommand{\ed}{\end{description}}
\newcommand{\bdm}{\begin{displaymath}}
\newcommand{\edm}{\end{displaymath}}

\newcommand{\gsim}{\mbox{$\:\stackrel{>}{_{\sim}}\:$} }

\newcommand{\hMpc}{{\ifmmode{h^{-1}{\rm Mpc}}\else{$h^{-1}$Mpc }\fi}}
\newcommand{\hGpc}{{\ifmmode{h^{-1}{\rm Gpc}}\else{$h^{-1}$Gpc }\fi}}
\newcommand{\hkpc}{{\ifmmode{h^{-1}{\rm kpc}}\else{$h^{-1}$kpc }\fi}}
\newcommand{\hMsun}{{\ifmmode{h^{-1}{\rm {M_{\odot}}}}\else{$h^{-1}{\rm{M_{\odot}}}$}\fi}}
\newcommand{\Msun}{{\ifmmode{{\rm {M_{\odot}}}}\else{${\rm{M_{\odot}}}$}\fi}}

\title{\boldmath Local gravitational redshifts can bias cosmological measurements}

\author[a,b]{Rados{\l}aw Wojtak,}
\author[c]{Tamara M. Davis,}
\author[b]{and Jophiel Wiis}

\affiliation[a]{Kavli Institute for Particle Astrophysics and Cosmology, Stanford University, SLAC National Accelerator Laboratory, Menlo Park, CA 94025}
\affiliation[b]{Dark Cosmology Centre, Niels Bohr Institute, University of Copenhagen, Denmark}
\affiliation[c]{School of Mathematics and Physics, University of Queensland, QLD 4072, Australia}

\emailAdd{wojtak@stanford.edu}

\abstract{
Measurements of cosmological parameters via the distance-redshift relation usually rely on models 
that assume a homogenous universe. It is commonly presumed that the large-scale structure evident 
in our Universe has a negligible impact on the measurement if distances probed in observations are 
sufficiently large (compared to the scale of inhomogeneities) and are averaged over different directions 
on the sky. This presumption does not hold when considering the effect of the gravitational redshift 
caused by our local gravitational potential, which alters light coming from all distances and directions 
in the same way. Despite its small magnitude, this local gravitational redshift gives rise to noticeable 
effects in cosmological inference using SN Ia data. Assuming conservative prior knowledge of the local 
potential given by sampling a range of gravitational potentials at locations of Milky-Way-like galaxies 
identified in cosmological simulations, we show that ignoring the gravitational redshift effect in a standard 
data analysis leads to an additional systematic error of $\sim1\%$ in the determination of density parameters 
and the dark energy equation of state. We conclude that our local gravitational field affects 
our cosmological inference at a level that is important in future observations aiming to achieve percent-level 
accuracy.
}

\begin{document}
\maketitle
\flushbottom

\section{Introduction}

When standing on the surface of the Earth, one's observations are strongly influenced by one's location.  Even such fundamental things as the strength of gravity, differ depending on whether you are on a mountain-top, in a valley, or sitting over an ore deposit.  To get an unbiased measurement, one needs to make measurements in many different places -- or take into account the local environment during the analysis. 

In cosmology, too, our local environment influences our view of the universe as a whole. The universe is arguably more homogeneous than the surface of the Earth, which is why the cosmological principle (that the universe is homogeneous and isotropic) has been a very successful basis of our studies of the expansion of the universe.  However, in this era of `precision cosmology' the inhomogeneities are no longer as negligible as they once were.  We are now at the stage where, to get accurate measurements of cosmological parameters, we need to take these inhomogeneities into account \citep[e.g.][]{Sinclair10,Davis11,Mar2013,Val2013,Ame2010,deLav2011,Bus2013}. 

Gravitational lensing and peculiar velocities are observational manifestations of large-scale structures in the universe. For cosmological purposes (as long as the distances being measured are larger than 
the local inhomogeneity), many of the effects 
will average out over many observations in many directions. 
For example peculiar velocities of distant galaxies will not strongly effect the cosmological measurements from supernovae, despite the fact that they are shifting the observed redshift from the true cosmological redshift, because this essentially adds random scatter, which averages out.  Similarly, lensing of supernova light will strongly magnify a few supernovae and slightly de-magnify most supernovae, because there is more volume in voids than clusters, but even though the scatter about the Hubble diagram is not gaussian, the mean remains unbiased \citep{Sar2008,Smith2014}.

The one effect that does not cancel, however, is the effect of our own location in this landscape of space.  Observations of our local universe suggest that we live in a slight underdensity, compared to the mean density of the universe, which extends up to 
$\sim100\hMpc$. The first observational indication of a local underdensity 
was shown in the data of supernovae Ia \citep[][]{Zeh1998,Jha2007}. This reasoning, 
however, happened to depend strongly on modeling the supernovae colors \citep{Con2007}. Many prevalent 
arguments for a local underdensity come from studies of galaxy counts 
\citep{Hua1997,Fri2003,Buss2004,Bal2008,Whi2014,Kee2013} and galaxy cluster counts \citep{Boe2015} in the local universe. 
If true, this would add a slight gravitational redshift to every redshift we measure, even those well outside our local density fluctuation.  It also means that peculiar velocities of local galaxies will tend to be away from us (being sucked out of our underdensity), thus increasing the apparent Hubble flow of our local universe.  The former of those effects -- the gravitational redshift of our own underdensity -- is the more important of the two for distant cosmological probes, because it effects every redshift we see from distant galaxies systematically.  Although the gravitational redshift we expect is very small ($\sim10^{-5}$), it can still have a noticeable effect on our cosmological parameter measurements because it is systematic.  

Gravitational redshift on cosmological scales was first detected in galaxy clusters \citep{Woj2011,Dom2012,Sad2015,Jim2015}. 
The redshift manifests itself as relative shifts of the galaxy velocity distributions at different distances from the cluster center \citep[for more details see][]{Cappi1995,Kim2004,Zha2013,Kaiser2013}. The measured gravitational redshift is consistent with typical depths of the gravitational potential well in cluster-mass dark matter haloes, i.e.\ approximately $4\times10^{-5}$ for haloes with virial masses of $10^{14}\Msun$. Comparable values of the gravitational redshift are expected from large-scale structures. The gravitational redshift on these scales is expected to give rise to asymmetric features in the cross-correlation function of massive galaxies with lower mass galaxies, which should be detectable in upcoming redshifts surveys \citep{Cro2013}.

The purpose of this paper is to estimate the magnitude of the gravitational redshift effect, and estimate the impact it has on our cosmological inferences.  We begin in Section~\ref{sect:gravz} by estimating the typical magnitude of gravitational redshift that light experiences as it travels from emitters to us, and assessing what part of that does not cancel out on average.  We consider light emitted from the mean gravitational potential of the universe, as well as light emitted from the mean gravitational potential of emitters, which are biased tracers of the potential (typically emitters sit in stronger gravitational wells because that is where galaxies form).

In Section~\ref{sect:effect} we assess the impact such a systematic shift in redshift would have on our cosmological inferences from supernova surveys.  We show how the impact changes as more flexibility is added to the models (e.g.\ allowing a varying equation of state).  Although the effect of a local gravitational redshift is small, it remains potentially important if we want to measure cosmological parameters to 1\% precision, and could cause false tensions between data sets.  We discuss these issues and conclude in Section~\ref{sect:conclusions}.

\section{Gravitational Redshift}\label{sect:gravz}

The gravitational redshift results from a difference in a gravitational potential between the point of light reception and emission. In the weak field limit, the gravitational 
redshift $z_{\rm g}$ is given by
\begin{equation}
z_{\rm g}=\frac{\phi_{\rm r}-\phi_{\rm e}}{c^{2}},
\label{redshift-general}
\end{equation}
where $\phi_{\rm r}$ and $\phi_{\rm e}$ is the gravitational potential at the points of light reception and emission, respectively. The gravitational 
redshift may take a positive or negative sign, depending on whether photons escape from a potential well or they are received 
therein. In the latter case, the gravitational 
redshift would be observed as a blueshift. We hereafter refer to both situations as the gravitational redshift and we differentiate 
both by sign: positive for redshift and negative for blueshift.

In cosmological context, the gravitational redshift is directly related to inhomogeneities in the matter distribution. The redshift 
can be calculated using equation (\ref{redshift-general}) with the peculiar gravitational potential given by the Poisson equation
\begin{equation}
\nabla^{2}\phi=4\pi Ga^{2}\bar{\rho}\delta(x,t),
\label{Poisson}
\end{equation}
where $a$ is the scale factor normalised to one at the present day, $\bar{\rho}$ is the mean background density (mass per physical volume, so $\bar{\rho}=\bar{\rho}_{0}a^{-3}$), the density contrast is $\delta=\rho/\bar{\rho}-1$, and the gradient is with respect to comoving coordinates \citep{Pee1980}.

Before we proceed with more detailed calculations, it is instructive  
to make a few simple estimates of the gravitational redshift for typical large-scale structures such as superclusters or cosmic voids. Approximating the matter distribution 
with a top-hat model, one can show that the peculiar gravitational potential is given by
\begin{equation}
\phi=\frac{G\Delta M}{R}=-\delta_{R}\Omega_{\rm m} H_{0}^{2}R^{2}/2,
\end{equation}
where $\Delta M$ is the total mass excess (decrement) with respect to the background density contribution, $R$ is the radius of a structure, $\delta_{R}$ is the total overdensity 
(the mean density contrast within $R$), $\Omega_{m}$ is the matter density parameter. We consider 
parameters characterizing two examples of observed large-scale structures: $\delta_{R}=0.8$ and $R=50$ $h^{-1}$ 
Mpc for the most massive superclusters similar to the Shapley Supercluster \citep{Mun2008},  $\delta_{R}=-0.3$ and 
$R=100$ $h^{-1}$ Mpc corresponding to large cosmic voids. This yields 
$\phi/c^{2}=-3\times10^{-5}$ and $\phi/c^{2}=5\times10^{-5}$ for the supercluster and the void, respectively. 
These are quite typical values of the gravitational redshift due to large-scale structures. Depending on the position 
of observer in the cosmic web, the gravitational redshift may be as large as $10^{-4}$ (or $30$ km/s, 
when expressed in terms of the approximate Doppler velocity $z_{\rm g}c$). It is interesting to notice that this 
value is comparable to the observed gravitational redshift due to deep potential wells in the most massive dark matter haloes 
associated with galaxy clusters \citep{Woj2011,Sad2015,Jim2015}. This means that our simple estimate of the gravitational 
redshift can be boosted by factor of 2 when considering light emitted from massive galaxy clusters embedded in large 
superclusters.

Contrary to the density field, the amplitude of the peculiar gravitational potential does not decay with increasing scale. 
This can be shown by considering the variance of matter density fluctuations -- and thus  potential fluctuations -- at large smoothing scales. In this regime, the power spectrum can be approximated by $P(k)\propto k^{n}$, where $n$ is the spectral index of the primordial matter density fluctuations. 
For a top-hat window, the rms of the mass fluctuations within spheres of radius  $R\propto M^{1/3}$ is 
given by $\sigma_M(R)\propto R^{-(3+n)/2}$ \citep{Pea1999}. 
Conversion into the rms of the potential fluctuations yields the following scaling relation: 
$\sigma_\phi(R)=G\sigma_M(R)M/R\propto R^{(1-n)/2}$.  For the Harrison-Zeldovich power spectrum ($n=1$), the rms of the potential fluctuations remains the same at all scales (in the limit of large scales) with $\sigma_\phi/c^{2}\approx 2\times 10^{-5}$ 
for $\Omega_{m}=0.3$ and the power spectrum normalised to $\sigma_{8}=0.8$ (approximately the measured rms of matter 
density fluctuations at a scale of $8\hMpc$). For a spectral index $n<1$, fluctuations of the gravitational potential start to grow at large scales and become a more dominant source of the potential compared to local structures at scales of $\sim10\hMpc$. This persistent lack of vanishing of large-scale Fourier modes is a dramatically different property compared to the density field. Arguably the most striking consequence of it is the fact that in terms of the peculiar potential, the Universe is inhomogenous at all scales. From a practical point of view, this property also shows that all calculations involving large-scale gravitational potential should include primarily large-scale Fourier modes. In particular, this implies the necessity of using cosmological simulations run in reasonably large boxes, with a side length of at least $\sim1\hGpc$.

\subsection{Gravitational redshift from cosmological simulations}

Exact theoretical calculations of the gravitational redshift should take into account non-linear evolution of  cosmic structures. The reason is twofold. First, one should expect non-negligible contribution to the 
potential from structures evolving in strictly non-linear regime, e.g. clusters in superclusters. Second, 
realistic predictions for observations should be based on a mock catalog of galaxies which define in 
a natural way a set of observed objects as well as locations of observers in the cosmic web.

\begin{figure}
\begin{center}
    \leavevmode
    \epsfxsize=14cm
    \epsfbox[100 65 1040 450]{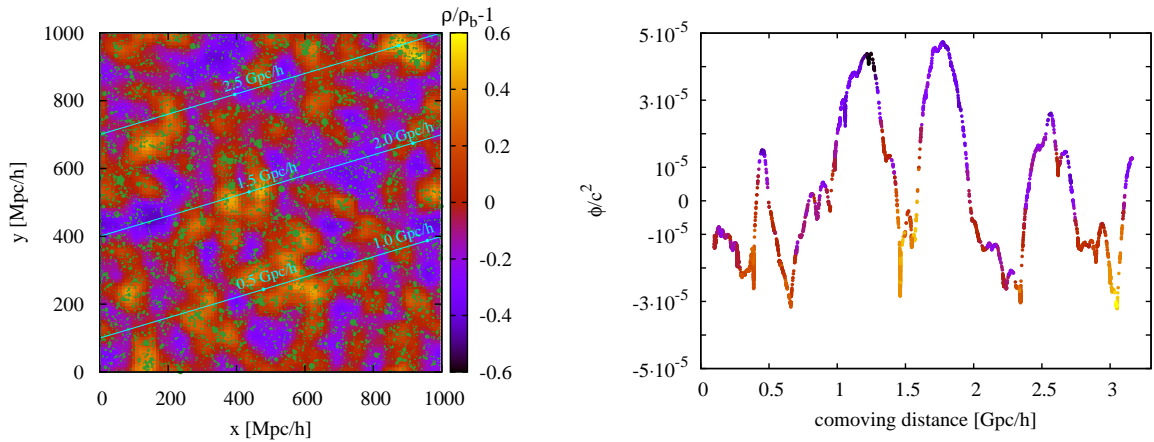}
\end{center}
\caption{A density contrast map from a MultiDark simulation smoothed at a scale of $50\hMpc$ (left panel); with the peculiar gravitational potential of dark matter particles along a beam-sized line crossing the simulation box (right panel). The color coding on both panels indicates consistently the smoothed density contrast. The green filled circles on the left panel show the positions of 
dark matter haloes along the sight-line chosen. Their sizes are proportional to the virial radii. Fluctuations of the 
gravitational potential reflects large-scale inhomogeneities of the matter distribution, 
whereas sharp minima coincide with massive dark matter haloes. Humps and troughs 
in the potential encompass regions dominated by gravitational collapse (superclusters) 
and evacuation of matter (voids), respectively.}
\label{line}
\end{figure}

For our calculation of the gravitational redshift, we employ a large-scale cosmological $N$-body simulation from 
the MultiDark data base.\footnote{The simulation is publicly available through the MultiDark database (http://www.multidark.org). 
See \citep{Rie2013} for all details of the database.} The simulation follows the evolution of $2048^{3}$ dark matter particles in a volume of $(1\hGpc)^{3}$. The particle mass is $8.7\times10^{8}\hMsun$, yielding a minimum resolved halo mass of around $10^{11}\hMsun$. The simulation adopted cosmological parameters determined from the fifth data release of WMAP satellite observations 
\citep{Kom2009}, i.e.\ $\Omega_{m}=0.27$, $\Omega_{\Lambda}=0.73$, $\sigma_{8}=0.82$.

As the first step of the calculation, we compute the peculiar gravitational potential at positions of all particles. The potential is computed with the Poisson solver built 
in the Gadget code \citep{Spr2005}. The right panel of Fig.~\ref{line} shows the gravitational potential of DM particles along a beam-size sight line crossing the simulation box at redshift $z=0$ (see the left panel). The color coding indicates the density contrast smoothed at a scale of $50\hMpc$. It is clearly readable that the potential profile is determined primarily by two effects: fluctuations due to large-scale structures at scales of $\geq 100\hMpc$ and local minima coinciding with the positions of massive dark matter haloes. As expected, minima of the potential occur at places 
undergoing the process of collapse (positive density contrast), whereas the maxima coincide with regions dominated by the process of matter evacuation (negative density contrast). The amplitude of the large-scale fluctuations is consistent with our simple estimates of the potential obtained for voids and superclusters with the matter distribution approximated by a top-hat function.

The gravitational redshift is a differential effect for which one needs to specify positions where light is emitted and received. 
It is natural to think that both places should be located in galaxies. In order to 
generate a catalog of galaxies we apply an HOD (Halo Occupation Distribution) model to the catalog of dark matter haloes 
found in the simulation. We use all distinct haloes (haloes which are not subhaloes of larger haloes) detected by 
the Bound-Density-Maxima halo finder \citep{Kly1997}. We assume that every halo has one central galaxy and 
a number of satellite galaxies. The satellite number for a given halo mass is drawn from a Poisson distribution whose 
mean is a power-law function of the halo mass. Parameters defining the mass dependance of the mean number of the satellites are taken from \citep{Zhe2005}. Once the galaxy catalog is generated, every galaxy is assigned the mean gravitational potential 
of all dark matter particles within the virial sphere of its host dark matter halo.

The gravitational redshift depends on the position of observer and selection of observed galaxies. The most straightforward way to quantify this effect is to consider a set of observers and a set of observed galaxies, and then 
to compute the distribution of mean (averaged over the same set of galaxies) gravitational redshifts as measured by 
these observers. In this approach, the gravitational redshift corresponds a difference between the gravitational 
potential at the observer's position and the mean potential at locations of all observed galaxies. Its distribution 
reflects directly the distribution of observers in the cosmic web (subject to a fixed observational selection). 
The same approach was recently used to study cosmic variance of the local determination of the Hubble constant 
\citep[][]{Woj2014}.

\begin{figure}
\begin{center}
    \leavevmode
    \epsfxsize=10cm
    \epsfbox[65 65 560 420]{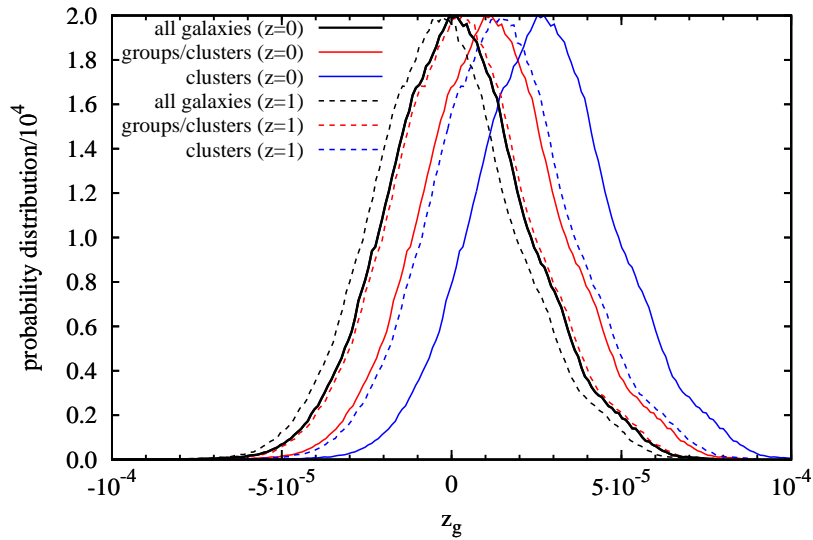}
\end{center}
\caption{Probability distributions of the gravitational redshift at positions of Milky-Way-like galaxies 
in a large-scale cosmological simulation at cosmological redshift $z=0$. The gravitational redshift is measured from 
light emitted from all galaxies formed in the simulation (black), galaxies in groups or clusters with a minimum host 
halo mass of $10^{13}\hMsun$ (red) or $10^{14}\hMsun$ (blue). The widths 
of the distributions represent a typical scatter in the gravitational potential due to locations of 
observers in the cosmic web, whereas gradual shifts of the distributions show effect of selecting galaxies 
in denser environments (solid red and blue lines) or at higher redshift $z=1$ (dashed lines). Positive (negative) 
values correspond to the actual redshift (blueshift) of observed spectra.}
\label{pdf-z0}
\end{figure}

We begin with a general setup for the calculation: observers located in Milky-Way-like galaxies and observing all galaxies found in the simulation at redshift $z=0$ what corresponds to a complete and shallow 
(no redshift evolution) galaxy survey. The Milky-Way galaxies are identified as the central galaxies 
of dark matter haloes with masses $(1-2.5)\times10^{12}\hMsun$, where the adopted mass range comes from 
recent observational constraints on the Milky-Way halo mass \citep[][]{Li2008}. The black line in 
Fig.~\ref{pdf-z0} shows the resulting distribution of the gravitational redshift (the same set of galaxies 
observed by observers located at different positions in the simulation box). The distribution is determined 
by the positions of observers in the cosmic web. Depending on whether observers are located in overdense 
or underdense regions, light from the observed galaxies becomes respectively blueshifted (negative values) 
or redshifted (positive values). The most likely value of the gravitational redshift is close to 0, while $95$ per cent of the allowed gravitational redshifts range between $\pm4\times10^{-5}$ ($z_{\rm g}c=\pm12$ km/s).

As a consequence of a flux limitation, most surveys tend to target more luminous galaxies which in turn populate overdense regions of the Universe.  Similarly, supernovae surveys may produce more supernova discoveries in denser fields on the sky. 
This kind of selection can be quite easily incorporated 
in our calculation by picking galaxies hosted by more massive haloes. The red and blue lines in Fig.~\ref{pdf-z0} 
show the distributions of the gravitational redshift given by the gravitational potential of galaxies residing in 
haloes with masses larger than $10^{13}\hMsun$ and $10^{14}\hMsun$, respectively (with the same set of 
observers defined by the Milky-Way-like galaxies). The two adopted minimum halo masses correspond to 
environments defined by selecting groups or clusters of galaxies. It is clearly visible from the plot that selecting 
galaxies in denser environments increases the probability of redshift against blueshift, i.e.\ a positive value of the 
gravitational redshift is more probable than negative.

\subsection{Time dependence of the gravitational redshift}

The above calculations are based on a $z=0$ snapshot, therefore they do not account for the 
time evolution of cosmic structures. A simple way to show this effect is to select galaxies from 
a higher-redshift snapshot. Here we repeat our calculations for redshift $z=1$ which is nearly 
an upper limit for the currently observed supernovae (there are far fewer 
at $z>1$). We consider the same set of observers as before, i.e. observers located in the Milky-Way-like 
galaxies at $z=0$, and analogous criteria for galaxy selection: all galaxies at $z=1$ and two subsets 
comprising galaxies in haloes with a minimum mass of $10^{13}\hMsun$ and $10^{14}\hMsun$. 
The dashed lines in Fig.~\ref{pdf-z0} show the resulting distributions of the gravitational redshift.

The apparent offsets between $z=0$ and $z=1$ distributions shown in Fig.~\ref{pdf-z0} 
result primarily from the growth of dark matter haloes. A time lag between formations of 
haloes with different masses \citep{Wech2002} explains dependence of the offset on the 
minimum halo mass used for galaxy selection: the offset becomes largest for the most 
recently formed cluster-sized haloes. Our calculations of the gravitational redshift distributions 
at $z=1$ automatically include evolution of larger-scale over- and under-densities which 
give rise to secondary anisotropies of cosmic microwave background photons propagating through 
these regions -- the integrated Sachs-Wolfe effect \citep[][]{Sachs1967}. The magnitude 
of the ISW redshift (due to time evolution of large-scale inhomogeneities) estimated 
in a standard $\Lambda$CDM cosmological model is $10^{-7}-10^{-6}$ \citep[][]{Nad2012,Gra2008}. 
This value is much smaller than the offsets between low- and high-redshift distributions shown 
in Fig.~\ref{pdf-z0}, therefore we conclude that the ISW effect is subdominant.

Bearing in mind that the scatter represented by the distributions in Fig.~\ref{pdf-z0} is predominantly 
determined by the distribution of observers in the cosmic web at $z=0$, it is not surprising that 
all gravitational redshift distributions have the same width. The offsets of the distributions due 
to redshift evolution are much smaller than their width. This means that in most practical 
applications the redshift evolution of the gravitational effect will be small.  Moreover, it is reassuring that the bias due to gravitational redshifts will be smaller when observing galaxies at higher-$z$ than at low-$z$ because structures were less evolved than they are now --- and thus density contrasts were not as significant. 

\subsection{Observers in extreme environments}

\begin{figure}
\begin{center}
    \leavevmode
    \epsfxsize=10cm
    \epsfbox[65 65 560 420]{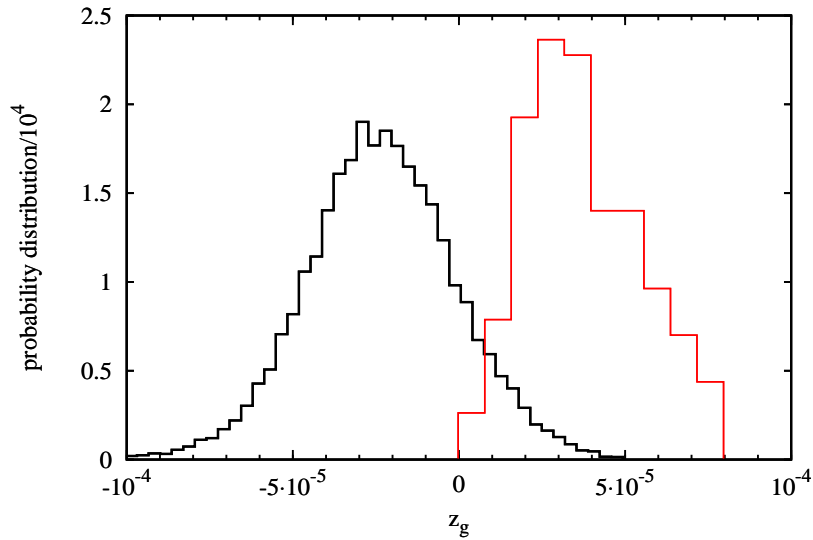}
\end{center}
\caption{Probability distribution of the gravitational redshift measured inside cosmic voids (red) or 
galaxy clusters (black) at redshift $z=0$. Observers in underdense environments tend to measure a 
positive signal (gravitational redshift), whereas those in galaxy clusters always 
observe a negative signal (gravitational blueshift).}
\label{pdf-z0-extr}
\end{figure}

The distributions of the gravitational redshifts shown in Fig.~\ref{pdf-z0} rely primarily on the assumed 
ensemble of observers. Although our choice of Milky-Way-like galaxies is fairly general, it seems to 
be instructive to explore other possibilities. In order to show a maximized effect of this assumption, 
we consider two extreme cases: observers located in galaxy clusters or in cosmic voids. Galaxy clusters 
are identified as dark matter haloes with a minimum mass of $10^{14}\hMsun$. Voids are found using 
ZOBOV void finder \citep{Ney2008} and observers are placed at locations of the minimum density 
smoothed at a scale of $10\hMpc$. We only consider the largest voids with a minimum effective radius 
of $100\hMpc$. Fig.~\ref{pdf-z0-extr} shows the resulting distributions of the gravitational redshift 
as measured from observations of all galaxies at $z=0$. As expected, observers in voids tend to measure a negative 
signal (gravitational blueshift), whereas those in galaxy clusters always observe gravitational redshift. 
Needless to say, both distributions sample respectively the lower and upper tails of the $z=0$ distributions 
shown in Fig.~\ref{pdf-z0}.

\section{Effect of gravitational redshift in analysis of SN Ia data}\label{sect:effect}

\begin{figure}
\begin{center}
    \leavevmode
    \epsfxsize=10cm
    \epsfbox[15 10 440 350]{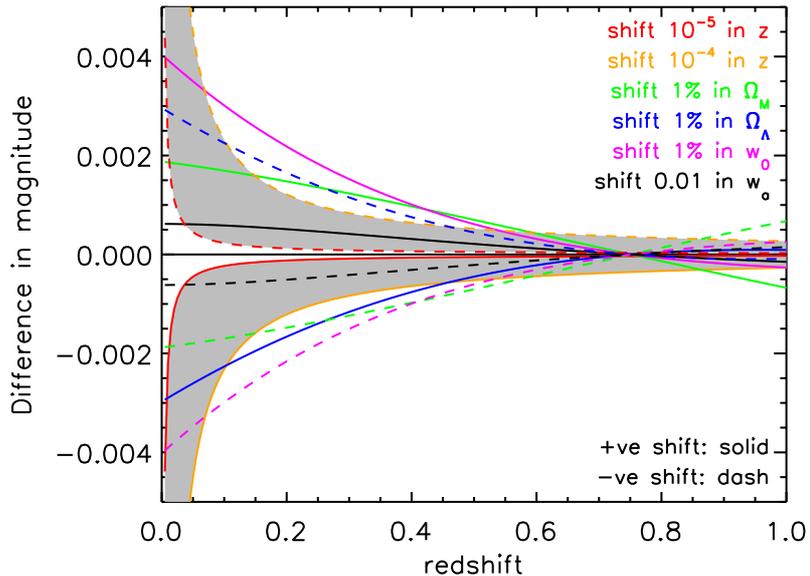}
\end{center}
\caption{
How a gravitational redshift changes the magnitude-redshift diagram, in comparison to how it is changed by shifts in cosmological parameters.  The horizontal axis is the observed redshift.  The shaded regions show how the magnitude-redshift relation is altered with respect to a fiducial $\Lambda$CDM model with $\Omega_{m}=0.3$ and $\Omega_{\Lambda}=0.7$, caused by the presence 
of the gravitational redshift ranging from $10^{-5}$ to $10^{-4}$ (solid lines for a positive redshift, dashed lines for a negative redshift). 
They are compared to analogous changes obtained by small perturbations of cosmological parameters in the fiducial model 
(with the amplitude of perturbations $0.01$ for $w_{a}$ and $1$ per cent of the fiducial values for the remaining 
parameters). Changes due to a gravitational redshift of $10^{-5}$--$10^{-4}$ 
can be partially mimicked by $\sim1\%$ perturbations in some cosmological parameters.  Supernova surveys with limited redshift range are particularly sensitive, because marginalising over absolute magnitude effectively allows an arbitrary vertical scaling.
}
\label{mag}
\end{figure}

Cosmological inference with SN data relies primarily on fitting the slope of 
the magnitude-redshift relation (or, equivalently, the distance modulus-redshift relation) and 
its dependance on cosmological redshift. The absolute normalization of this relation is degenerate 
with the Hubble constant and it is commonly treated as a nuisance parameter. 
Dependance of the apparent magnitude on cosmological redshift is strong enough, especially 
at small redshifts, to make the magnitude-redshift relation sensitive to quite small perturbations 
in redshift space. Even for perturbations as small as those given by the expected gravitational redshift 
due to large-scale structures, deviations in the distance modulus are large enough to have an impact 
on the measurement of cosmological parameters using SN data. This property is illustrated in 
Fig.~\ref{mag} which shows deviations of distance moduli (or apparent magnitude) from a fiducial 
$\Lambda$CDM model with $\Omega_{m}=0.3$ and $\Omega_{\Lambda}=0.7$, 
due to the presence of the local gravitational redshift. These deviations are compared to analogous 
changes in the magnitude caused by introducing small perturbations in cosmological parameters 
with respect to their fiducial values. We consider perturbations both in the density parameters, 
$\Omega_{m}$ and $\Omega_{\Lambda}$, and parameters of a time-dependent equation of state for 
dark energy, i.e. $w(a)=w_{0}+(1-a)w_{a}$ \citep[][]{Che2001,Linder2003}. The magnitude of perturbations 
is chosen to be $0.01$ for $w_{a}$ (its fiducial values is 0 when dark energy is a cosmological 
constant) and $1$ per cent of the fiducial 
values for all remaining parameters. The normalization of the magnitude-redshift relation is a nuisance 
parameter and cosmological fits would yield slightly different values for every set of perturbed data. Without 
loss of generality, we choose to normalise all models at an observed redshift of $0.75$.

Since the local slope of the magnitude-redshift relation gradually steepens with decreasing 
redshift, it is not surprising that perturbations in the apparent magnitude due to the local gravitational 
redshift diverge at small cosmological redshifts. This feature is clearly demonstrated in Fig.~\ref{mag}. It is also 
apparent from this figure that perturbations due to the local gravitational redshift can mimic small changes 
in cosmological parameters. This implies that the presence of the gravitational redshift is expected to affect 
determination of cosmological parameters with SN data.

Following the main idea illustrated in Fig.~\ref{mag}, here we quantify the impact of the gravitational redshift 
on cosmological parameters determined from SN data. The effect is measured by fitting cosmological 
parameters to a mock SN data set generated for a fiducial cosmological model and perturbed by 
the gravitational redshift. As a fiducial model we assume a flat $\Lambda$CDM model with 
$\Omega_{m}=0.3$ and $\Omega_{\Lambda}=0.7$. This model is used to calculate distances to 
and thus distance moduli of SNe. The mock data set comprises redshifts and errors in the apparent magnitude 
of all 557 SNe Ia from the Union2 compilation \citep{Ama2010}. We assume that these redshifts are 
exact cosmological redshifts which are then perturbed by applying a systematic shift due to the local 
gravitational redshift. As shown in the previous section (and Fig.~\ref{pdf-z0}), the time evolution of the gravitational redshift since 
redshift $z=1$ is small compared to a scatter due to the distribution of observers in the cosmic web. 
Therefore, for the purposes of this paper it is safe to assume that the perturbation due to gravitational redshift is independent of cosmological redshift. 

\begin{figure}
\begin{center}
    \leavevmode
    \epsfxsize=15.4cm
    \epsfbox[80 65 1120 420]{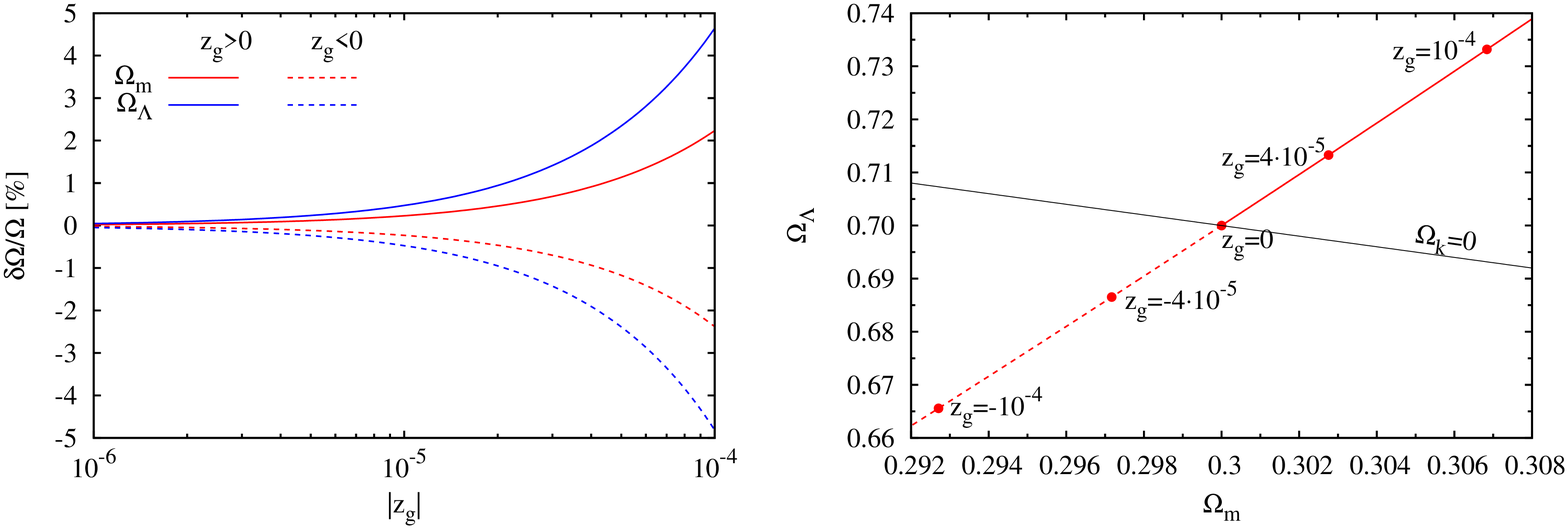}
\end{center}
\caption{Results of fitting a non-flat $\Lambda$CDM cosmological model to the mock SN data 
generated for a fiducial flat $\Lambda$CDM model with $\Omega_{m}=0.3$ and $\Omega_{\Lambda}=0.7$ 
and perturbed by the gravitational redshift due to the local gravitational potential. \textit{Left panel:} The relative 
deviation of the best fit parameters from the fiducial values as a function of the absolute magnitude 
of the gravitational redshift. The red and blue curves correspond to a positive and negative gravitational 
redshift. \textit{Right panel:} Migration of the best fit point on the plane of the density parameters. 
The black line shows flat $\Lambda$CDM models.
}
\label{OmOL}
\end{figure}

We fit a cosmological model to the mock SN data by minimizing the posterior probability marginalized 
over the normalization of the distance modulus-redshift relation. In this approach, the normalization 
is a nuisance parameter combining the absolute magnitude and the Hubble constant and no prior 
knowledge of its value is assumed in the analysis. As shown by \citep{Gol2001}, the logarithm of the 
posterior probability marginalized over the normalization parameter is given by
\begin{equation}
\chi^{2}(\theta)\propto\sum\frac{[\mu_{i}-\mu(\theta,z_{i})]^{2}}{\sigma_{i}^{2}}-
\frac{\Big(\sum\frac{\mu_{i}-\mu(\theta,z_{i})}{\sigma_{i}^{2}}\Big)^{2}}{\sum1/\sigma_{i}^{2}},
\end{equation}
where $\theta$ is a vector of cosmological parameters, $z_{i}$ are cosmological redshifts and 
$\sigma_{i}$ are the errors of the apparent magnitudes. The variable $\mu$ is a normalization-free 
distance modulus defined as
\begin{equation}
\mu=5\log_{10}(d_{L}),
\end{equation}
where $d_{L}$ is a Hubble-free luminosity distance given by
\begin{equation}\label{d_lum}
	d_{L}(z)= \propto \left\{
	\begin{array}{lll}
	 (1+z)\sin(I\sqrt{-\Omega_{k}})/\sqrt{-\Omega_{k}}, & \Omega_{k}<0\\
	 (1+z)I, & \Omega_{k}=0\\
	 (1+z)\sinh(I\sqrt{\Omega_{k}})/\sqrt{\Omega_{k}}, & \Omega_{k}>0,\\
\end{array} \right.
\end{equation}
with
\begin{equation}
\Omega_{k}=1-\Omega_{m}-\Omega_{\Lambda},
\end{equation}
\begin{equation}
I=\int_{0}^{z}\frac{\textrm{d}z'}{H(z')/H_{0}},
\end{equation}
\begin{equation}
H(z)/H_{0}=\sqrt{\Omega_{m}(1+z)^{3}+\Omega_{\Lambda}+\Omega_{k}(1+z)^{2}}.
\end{equation}
The $\mu$ variable is directly related to the apparent magnitude $m$ through the following relation
\begin{equation}
m=\mu+25+M+5\log_{10}(c/H_{0}),
\end{equation}
where $M$ is the absolute magnitude.

Fig.~\ref{OmOL} shows results of fitting $\Omega_{m}$ and $\Omega_{\Lambda}$ to the mock SN data, as a function 
of the assumed gravitational redshift. The left panel shows the relative deviations of the best fit parameters 
from the fiducial values and the right one shows a migration of the best fit point on the plane 
spanned by the two density parameters. It is clearly readable from this plot that the gravitational redshift has a noticeable effect on 
measuring cosmological parameters from SN data. A positive gravitational redshift (corresponding to a local underdensity) gives rise to an 
increase of the best fit density parameter at $z_{g}\leq5\times10^{-5}$ by up to $1\%$ for $\Omega_{m}$ and up to $3\%$ for 
$\Omega_{\Lambda}$. This leads effectively to a positive curvature with $0<\Omega_{k}\leq0.024$. 
For negative values of the gravitational redshift, both density parameters are biased low and the relative deviations 
are approximately symmetric to their counterparts with a positive gravitational redshift. The best fit parameters are 
distributed along a line which is oriented in a similar way as a degeneracy axis for a cosmological fit based 
on a non-flat $\Lambda$CDM model \citep[see e.g.][]{Ama2010}.

\begin{figure}
\begin{center}
    \leavevmode
    \epsfxsize=15.4cm
    \epsfbox[80 65 1120 420]{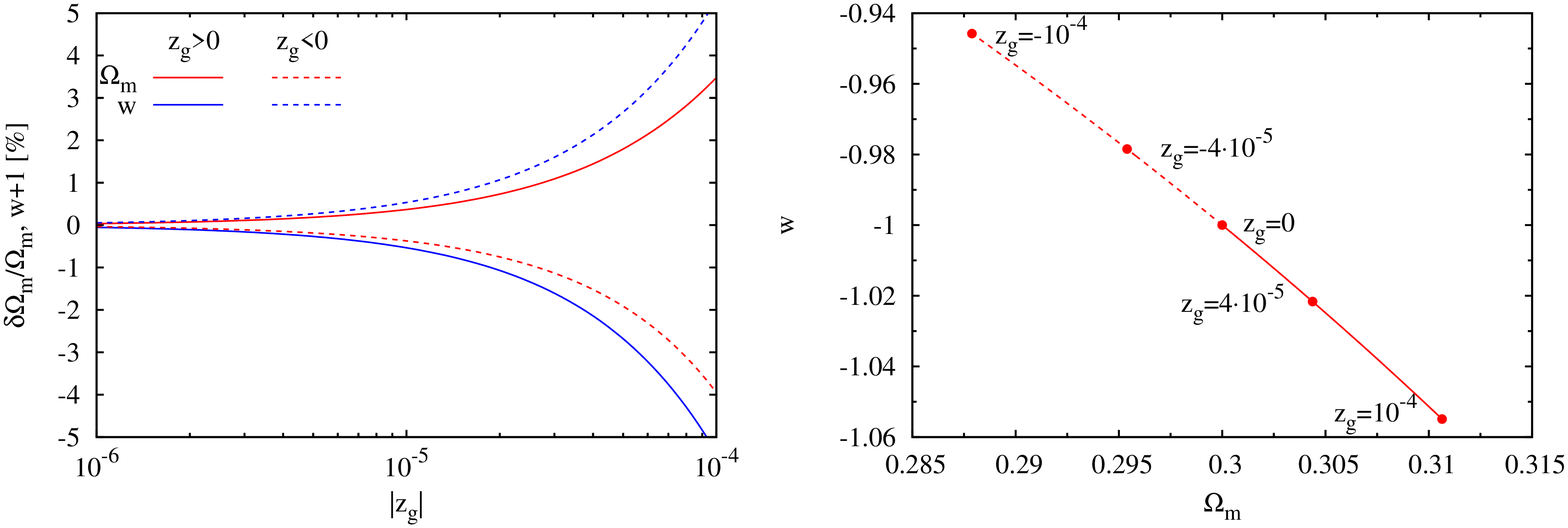}
\end{center}
\caption{Results of fitting a flat cosmological model with a free equation of state for dark energy, to the mock SN data 
generated for a fiducial flat $\Lambda$CDM model with $\Omega_{m}=0.3$ and $\Omega_{\Lambda}=0.7$ 
and perturbed by the gravitational redshift due to the local gravitational potential. \textit{Left panel:} The relative 
deviation of the best fit parameters from the fiducial values as a function of the absolute magnitude 
of the gravitational redshift.  The solid and dashed lines correspond to a positive and negative gravitation redshift, respectively.
\textit{Right panel:} Migration of the best fit point on the plane of $\Omega_{m}$ and $w$.
}
\label{Omw}
\end{figure}

Having shown that the gravitational redshift has a noticeable impact on the accuracy of measuring density 
of dark energy, it is instructive to consider whether constraints on the equation of state should be a matter 
of similar concern as well. Here we repeat cosmological fits assuming a flat cosmological model with a 
constant equation of state $w$.  The Hubble parameter in this case is given by:
\begin{equation}
H(z)/H_{0}=\sqrt{\Omega_{m}(1+z)^{3}+\Omega_{x}(1+z)^{3(1+w)}},
\end{equation}
where $\Omega_{x}$ is the density parameter for dark energy. Fig.~\ref{Omw} shows differences between the best 
fit $\Omega_{m}$ and $w$, and the fiducial values, i.e. $\Omega_{m}=0.3$ and $w=-1$. 
Compared to a cosmological fit with a $\Lambda$CDM parametrization, the presence of the gravitational 
redshift leads to opposite biases in $\Omega_{m}$ and $w$. A positive gravitational 
redshift increases the value of $\Omega_{m}$ and decreases the value of $w$. 
A change of the sign in the gravitational redshift reverses the signs of the biases keeping approximately 
the same absolute values. In general, effect of the gravitational redshift is stronger than for a $\Lambda$CDM fit. 
Similarly to a fit with a $\Lambda$CDM model, the line of best fit models in the parameter space 
resembles a degeneracy axis for a fit with a flat model and free equation of state $w$ \citep[see e.g.][]{Ama2010}.

\begin{figure}
\begin{center}
    \leavevmode
    \epsfxsize=15.4cm
    \epsfbox[80 65 1120 420]{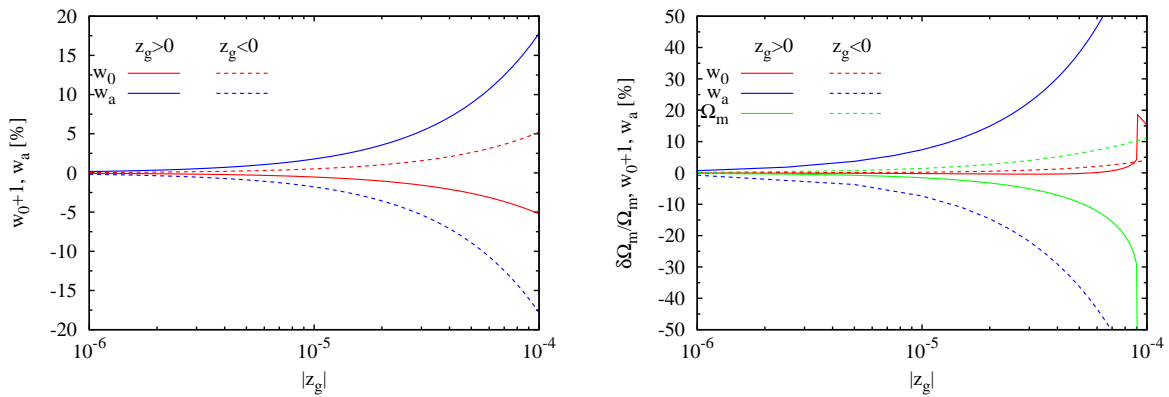}
\end{center}
\caption{Results of fitting a flat cosmological model with $\Omega_{\rm m}=0.3$ but with a free equation of state for dark energy that can vary linearly with scale factor, to the mock SN data 
generated for a fiducial flat $\Lambda$CDM model with $\Omega_{m}=0.3$ and $\Omega_{\Lambda}=0.7$ 
and perturbed by the gravitational redshift due to the local gravitational potential. The solid and dashed lines 
show the relative deviation of the best fit parameters from the fiducial values for a positive and negative 
redshift, respectively. \textit{Left panel:} Results for a model with free parameters of equation of state 
and fixed density parameters, i.e. $\Omega_{m}=0.3$ and $\Omega_{\Lambda}=0.7$. \textit{Right panel:} Results 
for a flat model with free parameters of equation of state. The peculiar behaviour of the $w_0$ fit at the far right 
occurs because the best fit $\Omega_{\rm m}$ hits zero -- which is the edge of the allowable parameter range.
}
\label{w0wa}
\end{figure}

In general, adding more degrees of freedom in cosmological model is expected to amplify deviations 
from the correct cosmological models. As an ultimate example we consider a cosmological model with a 
time-dependent equation of state for dark energy. We assume equation of state which is a linear function 
of the scale factor \citep[][]{Che2001,Linder2003},
\begin{equation}
w(a)=w_{0}+w_{a}(1-a),
\end{equation}
which yields the following redshift dependence of the Hubble parameter
\begin{eqnarray}
H(z)/H_{0} & = & \sqrt{\Omega_{m}(1+z)^{3}+\Omega_{x}f(z)}, \\
f(z) & = & (1+z)^{3(1+w_{0}+w_{a})}\exp[-3w_{a}z/(1+z)] \nonumber
\end{eqnarray}
As in the previous case, we assume a flat cosmological model. Fig.~\ref{w0wa} shows deviations of the best fit parameters 
obtained from fits assuming a fixed or varying $\Omega_{m}$. When keeping $\Omega_{m}$ fixed at its true value, deviations 
in $w_{a}$ are $3$ times larger than in $w_{0}$ which in turn exhibits nearly the same relative deviations as $w$ in a cosmological 
fit with a constant equation of state. A dramatic boost of the deviations is visible when one allows $\Omega_{m}$ to vary. 
In this case, best fit values of $w_{a}$ and $\Omega_{m}$ for gravitational redshift $z_{g}=\pm5\times 10^{-5}$ deviate by 
as much as $40\%$ and $\pm3\%$ from their true values. Stronger deviations at $z_{g}\geq7\times10^{-5}$ occur when 
$w_{0}+w_{a}$ becomes comparable to $0$. This part of parameter space violates early dark matter domination and it 
can be excluded by the cosmic microwave background (CMB) and baryon acoustic oscillations (BAO) 
data \citep{Ama2010}.   Note that if the gravitational redshift effect is important (if we are in a region of the 
universe experiencing  $z_g\gsim10^{-5}$) then an apparent tension between data sets would emerge until this 
effect was taken into account.

\subsection{Predictions for systematic errors}

There are two possible ways of dealing with the gravitational redshift effect when inferring cosmological 
parameters from SN data. Assuming prior knowledge of the exact value of the peculiar potential at our own location, 
one can attempt to correct the observed redshifts for the presence of the gravitational redshift. However, a plausible 
estimate of the local gravitational potential and thus the gravitational redshift is quite imprecise, because it can only 
be based on a model of the local density fluctuations and thus it cannot account for a non-negligible contribution 
from large-scale Fourier modes. Lack of precise observational constraints on the gravitational redshift leads to a second possibility 
in which the gravitational redshift becomes a source of additional systematic error. In this approach, we accept 
the fact that the gravitational redshift effect puts limits on the accuracy of cosmological inference using SN data, 
regardless of amount and quality of the data.  Note that this will also affect other distance probes such as baryon acoustic oscillations (BAO).  However in the case of BAO the existence of a calibrated standard ruler length (from the CMB) makes the fit more robust because it restricts the overall magnitude shift and thus removes flexibility from the fit.

\begin{figure}
\begin{center}
    \leavevmode
    \epsfxsize=10cm
    \epsfbox[65 45 560 408]{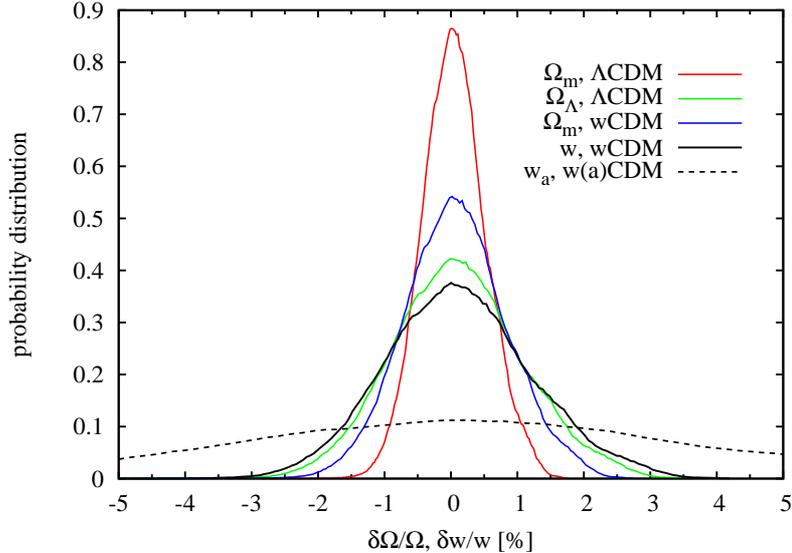}
\end{center}
\caption{Systematic errors in a measurement of cosmological parameters using SN data, 
due to the presence of an unconstrained gravitational redshift.
The curves show the distributions of differences between the best fit cosmological parameters and 
their fiducial values assumed for the mock SN data. The distributions reflect positions of the Milky-Way-like 
galaxies in the cosmic web. Cosmological 
parameters are obtained from two-parameter fits in a non-flat $\Lambda$CDM cosmological model with 
free $\Omega_{m}$ and $\Omega_{\Lambda}$, a flat $w$CDM model with free $\Omega_{m}$ and equation of state $w$, or 
a flat $w(a)$CDM model with free $w_{0}$ and $w_{a}$.
}
\label{sys-error}
\end{figure}

The systematic errors due to the presence of the gravitational redshift can be calculated by combining results 
from fitting the mock SN data with the expected distribution of the gravitational redshifts as measured by 
observers located in the Milky-Way-like galaxies. Since the redshift evolution of the mean potential is much 
smaller compared to a scatter due to locations of observers in the cosmic web, we only 
consider the gravitational redshift distribution calculated at cosmological redshift $z=0$. We also conservatively 
assume that the whole population of galaxies found in the simulation is possibly the least biased representation of a population of the SN Ia host galaxies (see the black curve in Fig.~\ref{pdf-z0}).

Fig.~\ref{sys-error} shows the distribution of differences between the best fit cosmological parameters and the fiducial values 
assumed for the SN mock data. Different kinds of lines show results for a non-flat $\Lambda$CDM model 
($\Omega_{m}$ and $\Omega_{\Lambda}$), a flat $w$CDM model with a free equation of state for 
dark energy ($\Omega_{m}$ and $w$), and  a flat $w_{a}$CDM model with equation of state linearly dependent on 
the scale factor ($w_{0}$ and $w_{a}$). The standard deviations of the distributions shall 
be interpreted as systematic errors of the measurement due to the presence of an unconstrained gravitational 
redshift related to the local gravitational potential. Relative systematic errors amount to $1\%$ and $0.5\%$ for 
$\Omega_{\Lambda}$ and $\Omega_{m}$ when fitting a non-flat $\Lambda$CDM cosmological model. The corresponding 
error in $\Omega_{k}$ is 0.01. When fitting a flat cosmological model with a free equation of state for dark energy, 
the expected systematic errors are $0.8\%$ for $\Omega_{m}$ and $1.1\%$  for $w$.  Finally, fitting a flat 
model with a redshift-dependent equation of state yields systematic error of $3.7\%$ in $w_{a}$. This uncertainty 
is boosted by a factor of $4$  when allowing $\Omega_{m}$ to vary.

While the effect of this local-density bias on the final cosmological model can be reduced significantly by combining 
supernova data with other cosmological probes, the bias will nevertheless cause a tension between data sets that 
will worsen as the precision of the measurements improves.  These tensions mean more complex models can falsely 
improve on the standard model by relieving the tension. In other words, biases in any data set can cause one to falsely 
conclude a more complex model is necessary to explain the data.  All biases should therefore be eliminated where possible.

\section{Discussion and conclusions}\label{sect:conclusions}

The gravitational redshift due to our own location in the large scale gravitational potential field becomes relevant 
when using SN data for determination of cosmological parameters to 1\% precision. The effect of the gravitational redshift 
on the best fit models is particularly amplified in extensions of a standard $\Lambda$CDM cosmological model 
such as models invoking dark energy with a redshift-dependent equation of state. These models are of particular 
attention from the point of view of future dark energy surveys whose goal is to discriminate between a cosmological 
constant and other forms of dark energy. Our findings imply that neglecting gravitational redshift in a cosmological analysis 
of SN data leads to an additional systematic uncertainty of 4\% in $w_{a}$ assuming an unbiased prior on $\Omega_{m}$, 
and 16\% when allowing $\Omega_{m}$ to vary.

A cosmological fit with SN data can in principle be corrected for the gravitational redshift effect. This correction would require 
precise measurement of the gravitational redshift at our location. However, current constraints on the matter 
distribution in the local universe are not accurate enough, so the only means to account for the gravitational redshift 
is to consider an additional systematic error caused by lack of prior knowledge of this redshift. Calculations of these 
systematic errors presented in this work are based on a matter distribution given by a standard $\Lambda$CDM model. 
This kind of analysis is circular in the sense that results are correct if $\Lambda$CDM model is valid. It is a minimum 
approach in which we show that the presence of the gravitational redshift permitted by a standard cosmological 
model can mimic certain modifications of the model itself. However, one should bear in mind that other matter 
distributions for other cosmological models are expected to result in different (and potentially larger) estimates 
of the systematic errors.

The presence of a gravitational redshift in SN data alters cosmological fits through an alteration in the distance modulus vs redshift relation (see Fig.~\ref{mag}).  This issue is worse for supernovae than BAO because for supernovae  the combination of absolute magnitude and Hubble parameter is not accurately known, effectively introducing an arbitrary additive scaling to the distance modulus that we need to marginalise over (in Fig.~\ref{mag} we arbitrarily normalised the theory curves to $z=0.75$). The BAO length scale 
can be calibrated to the CMB (or theory), reducing the flexibility in the fit and preventing a redshift shift 
from mimicking a cosmological parameter shift. Also, since BAO are a standard ruler while supernovae are a standard candle, the errors due to an undiagnosed local density fluctuation will differ between the two probes. Essentially this is because the angular diameter and luminosity distances differ by a factor of $(1+z)^2$, the former being proper distance divided by $(1+z)$ and the latter being proper distance multiplied by $(1+z)$. While the ratio of the luminosity distance to the angular diameter distance must equal to $(1+z)^2$ 
in any metric theory of gravity, in practice this {\em distance duality} relation can be broken because during the analysis 
the redshift has been mis-assumed to be entirely cosmological.

Finally, what we have shown is that small systematic shifts in redshift can have a large impact on cosmological results. 
Thus we should take extreme care when calibrating and fitting for the redshifts of supernovae to ensure no observational effects bias 
our cosmological results. For example, a fraction of a pixel systematic offset in the spectrograph wavelength calibration, an incorrect heliocentric correction, or an error in conversion to the CMB frame, could easily cause a shift larger than the one due to our local gravitational potential.  If that shift is systematic across the entire data set, it would also bias cosmological results in much the same 
way as a gravitational redshift.

\acknowledgments
We thank Krzysztof Bolejko, David Parkinson and Risa Wechsler for useful discussions.  
We also thank the anonymous referee for insightful comments. TMD acknowledges the support of the 
Australian Research Council through a Future Fellowship, grant FT100100595. RW acknowledges support through the Porat 
Postdoctoral Fellowship. The Dark Cosmology Centre is funded by the Danish National Research Foundation.

\bibliographystyle{jhep}
\bibliography{bibGravz}

\providecommand{\href}[2]{#2}\begingroup\raggedright\begin{thebibliography}{10}

\bibitem{Sinclair10}
B.~{Sinclair}, T.~M. {Davis}, and T.~{Haugb{\o}lle}, {\it {Residual
  Hubble-bubble Effects On Supernova Cosmology}},  {\em \apj} {\bf 718} (Aug.,
  2010) 1445--1455, [\href{http://arxiv.org/abs/1006.0911}{{\tt
  arXiv:1006.0911}}].

\bibitem{Davis11}
T.~M. {Davis}, L.~{Hui}, J.~A. {Frieman}, T.~{Haugb{\o}lle}, R.~{Kessler},
  B.~{Sinclair}, J.~{Sollerman}, B.~{Bassett}, J.~{Marriner},
  E.~{M{\"o}rtsell}, R.~C. {Nichol}, M.~W. {Richmond}, M.~{Sako}, D.~P.
  {Schneider}, and M.~{Smith}, {\it {The Effect of Peculiar Velocities on
  Supernova Cosmology}},  {\em \apj} {\bf 741} (Nov., 2011) 67,
  [\href{http://arxiv.org/abs/1012.2912}{{\tt arXiv:1012.2912}}].

\bibitem{Mar2013}
V.~{Marra}, M.~{P{\"a}{\"a}kk{\"o}nen}, and W.~{Valkenburg}, {\it {Uncertainty
  on w from large-scale structure}},  {\em \mnras} {\bf 431} (May, 2013)
  1891--1902, [\href{http://arxiv.org/abs/1203.2180}{{\tt arXiv:1203.2180}}].

\bibitem{Val2013}
W.~{Valkenburg}, M.~{Kunz}, and V.~{Marra}, {\it {Intrinsic uncertainty on the
  nature of dark energy}},  {\em Physics of the Dark Universe} {\bf 2} (Dec.,
  2013) 219--223, [\href{http://arxiv.org/abs/1302.6588}{{\tt
  arXiv:1302.6588}}].

\bibitem{Ame2010}
L.~{Amendola}, K.~{Kainulainen}, V.~{Marra}, and M.~{Quartin}, {\it
  {Large-Scale Inhomogeneities May Improve the Cosmic Concordance of
  Supernovae}},  {\em Physical Review Letters} {\bf 105} (Sept., 2010) 121302,
  [\href{http://arxiv.org/abs/1002.1232}{{\tt arXiv:1002.1232}}].

\bibitem{deLav2011}
A.~{de Lavallaz} and M.~{Fairbairn}, {\it {Effects of voids on the
  reconstruction of the equation of state of dark energy}},  {\em \prd} {\bf
  84} (Oct., 2011) 083005, [\href{http://arxiv.org/abs/1106.1611}{{\tt
  arXiv:1106.1611}}].

\bibitem{Bus2013}
V.~C. {Busti}, R.~F.~L. {Holanda}, and C.~{Clarkson}, {\it {Supernovae as
  probes of cosmic parameters: estimating the bias from under-dense lines of
  sight}},  {\em \jcap} {\bf 11} (Nov., 2013) 20,
  [\href{http://arxiv.org/abs/1309.6540}{{\tt arXiv:1309.6540}}].

\bibitem{Sar2008}
D.~{Sarkar}, A.~{Amblard}, D.~E. {Holz}, and A.~{Cooray}, {\it {Lensing and
  Supernovae: Quantifying the Bias on the Dark Energy Equation of State}},
  {\em \apj} {\bf 678} (May, 2008) 1--5,
  [\href{http://arxiv.org/abs/0710.4143}{{\tt arXiv:0710.4143}}].

\bibitem{Smith2014}
M.~{Smith}, D.~J. {Bacon}, R.~C. {Nichol}, H.~{Campbell}, C.~{Clarkson},
  R.~{Maartens}, C.~B. {D'Andrea}, B.~A. {Bassett}, D.~{Cinabro}, D.~A.
  {Finley}, J.~A. {Frieman}, L.~{Galbany}, P.~M. {Garnavich}, M.~D. {Olmstead},
  D.~P. {Schneider}, C.~{Shapiro}, and J.~{Sollerman}, {\it {The Effect of Weak
  Lensing on Distance Estimates from Supernovae}},  {\em \apj} {\bf 780} (Jan.,
  2014) 24, [\href{http://arxiv.org/abs/1307.2566}{{\tt arXiv:1307.2566}}].

\bibitem{Zeh1998}
I.~{Zehavi}, A.~G. {Riess}, R.~P. {Kirshner}, and A.~{Dekel}, {\it {A Local
  Hubble Bubble from Type IA Supernovae?}},  {\em \apj} {\bf 503} (Aug., 1998)
  483--491, [\href{http://arxiv.org/abs/astro-ph/9802252}{{\tt
  astro-ph/9802252}}].

\bibitem{Jha2007}
S.~{Jha}, A.~G. {Riess}, and R.~P. {Kirshner}, {\it {Improved Distances to Type
  Ia Supernovae with Multicolor Light-Curve Shapes: MLCS2k2}},  {\em \apj} {\bf
  659} (Apr., 2007) 122--148,
  [\href{http://arxiv.org/abs/astro-ph/0612666}{{\tt astro-ph/0612666}}].

\bibitem{Con2007}
A.~{Conley}, R.~G. {Carlberg}, J.~{Guy}, D.~A. {Howell}, S.~{Jha}, A.~G.
  {Riess}, and M.~{Sullivan}, {\it {Is There Evidence for a Hubble Bubble? The
  Nature of Type Ia Supernova Colors and Dust in External Galaxies}},  {\em
  \apjl} {\bf 664} (July, 2007) L13--L16,
  [\href{http://arxiv.org/abs/0705.0367}{{\tt arXiv:0705.0367}}].

\bibitem{Hua1997}
J.-S. {Huang}, L.~L. {Cowie}, J.~P. {Gardner}, E.~M. {Hu}, A.~{Songaila},
  {Wainscoat}, and {R.~J.}, {\it {The Hawaii K-Band Galaxy Survey. II. Bright
  K-Band Imaging}},  {\em \apj} {\bf 476} (Feb., 1997) 12--21,
  [\href{http://arxiv.org/abs/astro-ph/9610084}{{\tt astro-ph/9610084}}].

\bibitem{Fri2003}
W.~J. {Frith}, G.~S. {Busswell}, R.~{Fong}, N.~{Metcalfe}, and T.~{Shanks},
  {\it {The local hole in the galaxy distribution: evidence from 2MASS}},  {\em
  \mnras} {\bf 345} (Nov., 2003) 1049--1056,
  [\href{http://arxiv.org/abs/astro-ph/0302331}{{\tt astro-ph/0302331}}].

\bibitem{Buss2004}
G.~S. {Busswell}, T.~{Shanks}, W.~J. {Frith}, P.~J. {Outram}, N.~{Metcalfe},
  and R.~{Fong}, {\it {The local hole in the galaxy distribution: new optical
  evidence}},  {\em \mnras} {\bf 354} (Nov., 2004) 991--1004,
  [\href{http://arxiv.org/abs/astro-ph/0302330}{{\tt astro-ph/0302330}}].

\bibitem{Bal2008}
I.~K. {Baldry}, K.~{Glazebrook}, and S.~P. {Driver}, {\it {On the galaxy
  stellar mass function, the mass-metallicity relation and the implied baryonic
  mass function}},  {\em \mnras} {\bf 388} (Aug., 2008) 945--959,
  [\href{http://arxiv.org/abs/0804.2892}{{\tt arXiv:0804.2892}}].

\bibitem{Whi2014}
J.~R. {Whitbourn} and T.~{Shanks}, {\it {The local hole revealed by galaxy
  counts and redshifts}},  {\em \mnras} {\bf 437} (Jan., 2014) 2146--2162,
  [\href{http://arxiv.org/abs/1307.4405}{{\tt arXiv:1307.4405}}].

\bibitem{Kee2013}
R.~C. {Keenan}, A.~J. {Barger}, and L.~L. {Cowie}, {\it {Evidence for a \~{}300
  Megaparsec Scale Under-density in the Local Galaxy Distribution}},  {\em
  \apj} {\bf 775} (Sept., 2013) 62, [\href{http://arxiv.org/abs/1304.2884}{{\tt
  arXiv:1304.2884}}].

\bibitem{Boe2015}
H.~{B{\"o}hringer}, G.~{Chon}, M.~{Bristow}, and C.~A. {Collins}, {\it {The
  extended ROSAT-ESO Flux-Limited X-ray Galaxy Cluster Survey (REFLEX II). V.
  Exploring a local underdensity in the southern sky}},  {\em \aap} {\bf 574}
  (Feb., 2015) A26, [\href{http://arxiv.org/abs/1410.2172}{{\tt
  arXiv:1410.2172}}].

\bibitem{Woj2011}
R.~{Wojtak}, S.~H. {Hansen}, and J.~{Hjorth}, {\it {Gravitational redshift of
  galaxies in clusters as predicted by general relativity}},  {\em \nat} {\bf
  477} (Sept., 2011) 567--569, [\href{http://arxiv.org/abs/1109.6571}{{\tt
  arXiv:1109.6571}}].

\bibitem{Dom2012}
M.~J.~d.~L. {Dom{\'{\i}}nguez Romero}, D.~{Garc{\'{\i}}a Lambas}, and
  H.~{Muriel}, {\it {An improved method for the identification of galaxy
  systems: measuring the gravitational redshift by dark matter haloes}},  {\em
  \mnras} {\bf 427} (Nov., 2012) L6--L10.

\bibitem{Sad2015}
I.~{Sadeh}, L.~L. {Feng}, and O.~{Lahav}, {\it {Gravitational Redshift of
  Galaxies in Clusters from the Sloan Digital Sky Survey and the Baryon
  Oscillation Spectroscopic Survey}},  {\em Physical Review Letters} {\bf 114}
  (Feb., 2015) 071103, [\href{http://arxiv.org/abs/1410.5262}{{\tt
  arXiv:1410.5262}}].

\bibitem{Jim2015}
P.~{Jimeno}, T.~{Broadhurst}, J.~{Coupon}, K.~{Umetsu}, and R.~{Lazkoz}, {\it
  {Comparing gravitational redshifts of SDSS galaxy clusters with the magnified
  redshift enhancement of background BOSS galaxies}},  {\em \mnras} {\bf 448}
  (Apr., 2015) 1999--2012, [\href{http://arxiv.org/abs/1410.6050}{{\tt
  arXiv:1410.6050}}].

\bibitem{Cappi1995}
A.~{Cappi}, {\it {Gravitational redshift in galaxy clusters.}},  {\em \aap}
  {\bf 301} (Sept., 1995) 6.

\bibitem{Kim2004}
Y.-R. {Kim} and R.~A.~C. {Croft}, {\it {Gravitational Redshifts in Simulated
  Galaxy Clusters}},  {\em \apj} {\bf 607} (May, 2004) 164--174,
  [\href{http://arxiv.org/abs/astro-ph/0402047}{{\tt astro-ph/0402047}}].

\bibitem{Zha2013}
H.~{Zhao}, J.~A. {Peacock}, and B.~{Li}, {\it {Testing gravity theories via
  transverse Doppler and gravitational redshifts in galaxy clusters}},  {\em
  \prd} {\bf 88} (Aug., 2013) 043013,
  [\href{http://arxiv.org/abs/1206.5032}{{\tt arXiv:1206.5032}}].

\bibitem{Kaiser2013}
N.~{Kaiser}, {\it {Measuring gravitational redshifts in galaxy clusters}},
  {\em \mnras} {\bf 435} (Oct., 2013) 1278--1286,
  [\href{http://arxiv.org/abs/1303.3663}{{\tt arXiv:1303.3663}}].

\bibitem{Cro2013}
R.~A.~C. {Croft}, {\it {Gravitational redshifts from large-scale structure}},
  {\em \mnras} {\bf 434} (Oct., 2013) 3008--3017,
  [\href{http://arxiv.org/abs/1304.4124}{{\tt arXiv:1304.4124}}].

\bibitem{Pee1980}
P.~J.~E. {Peebles}, {\em {The large-scale structure of the universe}}.
\newblock Princeton University Press, 1980.

\bibitem{Mun2008}
J.~A. {Mu{\~n}oz} and A.~{Loeb}, {\it {The density contrast of the Shapley
  supercluster}},  {\em \mnras} {\bf 391} (Dec., 2008) 1341--1349,
  [\href{http://arxiv.org/abs/0805.0596}{{\tt arXiv:0805.0596}}].

\bibitem{Pea1999}
J.~A. {Peacock}, {\em {Cosmological Physics}}.
\newblock Cambridge University Press, Jan., 1999.

\bibitem{Rie2013}
K.~{Riebe}, A.~M. {Partl}, H.~{Enke}, J.~{Forero-Romero}, S.~{Gottl{\"o}ber},
  A.~{Klypin}, G.~{Lemson}, F.~{Prada}, J.~R. {Primack}, M.~{Steinmetz}, and
  V.~{Turchaninov}, {\it {The MultiDark Database: Release of the Bolshoi and
  MultiDark cosmological simulations}},  {\em Astronomische Nachrichten} {\bf
  334} (Aug., 2013) 691--708.

\bibitem{Kom2009}
E.~{Komatsu}, J.~{Dunkley}, M.~R. {Nolta}, C.~L. {Bennett}, B.~{Gold},
  G.~{Hinshaw}, N.~{Jarosik}, D.~{Larson}, M.~{Limon}, L.~{Page}, D.~N.
  {Spergel}, M.~{Halpern}, R.~S. {Hill}, A.~{Kogut}, S.~S. {Meyer}, G.~S.
  {Tucker}, J.~L. {Weiland}, E.~{Wollack}, and E.~L. {Wright}, {\it {Five-Year
  Wilkinson Microwave Anisotropy Probe Observations: Cosmological
  Interpretation}},  {\em \apjs} {\bf 180} (Feb., 2009) 330--376,
  [\href{http://arxiv.org/abs/0803.0547}{{\tt arXiv:0803.0547}}].

\bibitem{Spr2005}
V.~{Springel}, {\it {The cosmological simulation code GADGET-2}},  {\em \mnras}
  {\bf 364} (Dec., 2005) 1105--1134,
  [\href{http://arxiv.org/abs/astro-ph/0505010}{{\tt astro-ph/0505010}}].

\bibitem{Kly1997}
A.~{Klypin} and J.~{Holtzman}, {\it {Particle-Mesh code for cosmological
  simulations}},  {\em ArXiv Astrophysics e-prints} (Dec., 1997)
  [\href{http://arxiv.org/abs/astro-ph/9712217}{{\tt astro-ph/9712217}}].

\bibitem{Zhe2005}
Z.~{Zheng}, A.~A. {Berlind}, D.~H. {Weinberg}, A.~J. {Benson}, C.~M. {Baugh},
  S.~{Cole}, R.~{Dav{\'e}}, C.~S. {Frenk}, N.~{Katz}, and C.~G. {Lacey}, {\it
  {Theoretical Models of the Halo Occupation Distribution: Separating Central
  and Satellite Galaxies}},  {\em \apj} {\bf 633} (Nov., 2005) 791--809,
  [\href{http://arxiv.org/abs/astro-ph/0408564}{{\tt astro-ph/0408564}}].

\bibitem{Woj2014}
R.~{Wojtak}, A.~{Knebe}, W.~A. {Watson}, I.~T. {Iliev}, S.~{He{\ss}},
  D.~{Rapetti}, G.~{Yepes}, and S.~{Gottl{\"o}ber}, {\it {Cosmic variance of
  the local Hubble flow in large-scale cosmological simulations}},  {\em
  \mnras} {\bf 438} (Feb., 2014) 1805--1812,
  [\href{http://arxiv.org/abs/1312.0276}{{\tt arXiv:1312.0276}}].

\bibitem{Li2008}
Y.-S. {Li} and S.~D.~M. {White}, {\it {Masses for the Local Group and the Milky
  Way}},  {\em \mnras} {\bf 384} (Mar., 2008) 1459--1468,
  [\href{http://arxiv.org/abs/0710.3740}{{\tt arXiv:0710.3740}}].

\bibitem{Wech2002}
R.~H. {Wechsler}, J.~S. {Bullock}, J.~R. {Primack}, A.~V. {Kravtsov}, and
  A.~{Dekel}, {\it {Concentrations of Dark Halos from Their Assembly
  Histories}},  {\em \apj} {\bf 568} (Mar., 2002) 52--70,
  [\href{http://arxiv.org/abs/astro-ph/0108151}{{\tt astro-ph/0108151}}].

\bibitem{Sachs1967}
R.~K. {Sachs} and A.~M. {Wolfe}, {\it {Perturbations of a Cosmological Model
  and Angular Variations of the Microwave Background}},  {\em \apj} {\bf 147}
  (Jan., 1967) 73.

\bibitem{Nad2012}
S.~{Nadathur}, S.~{Hotchkiss}, and S.~{Sarkar}, {\it {The integrated
  Sachs-Wolfe imprint of cosmic superstructures: a problem for
  {$\Lambda$}CDM}},  {\em \jcap} {\bf 6} (June, 2012) 42,
  [\href{http://arxiv.org/abs/1109.4126}{{\tt arXiv:1109.4126}}].

\bibitem{Gra2008}
B.~R. {Granett}, M.~C. {Neyrinck}, and I.~{Szapudi}, {\it {An Imprint of
  Superstructures on the Microwave Background due to the Integrated Sachs-Wolfe
  Effect}},  {\em \apjl} {\bf 683} (Aug., 2008) L99--L102,
  [\href{http://arxiv.org/abs/0805.3695}{{\tt arXiv:0805.3695}}].

\bibitem{Ney2008}
M.~C. {Neyrinck}, {\it {ZOBOV: a parameter-free void-finding algorithm}},  {\em
  \mnras} {\bf 386} (June, 2008) 2101--2109,
  [\href{http://arxiv.org/abs/0712.3049}{{\tt arXiv:0712.3049}}].

\bibitem{Che2001}
M.~{Chevallier} and D.~{Polarski}, {\it {Accelerating Universes with Scaling
  Dark Matter}},  {\em International Journal of Modern Physics D} {\bf 10}
  (2001) 213--223, [\href{http://arxiv.org/abs/gr-qc/0009008}{{\tt
  gr-qc/0009008}}].

\bibitem{Linder2003}
E.~V. {Linder}, {\it {Exploring the Expansion History of the Universe}},  {\em
  Physical Review Letters} {\bf 90} (Mar., 2003) 091301,
  [\href{http://arxiv.org/abs/astro-ph/0208512}{{\tt astro-ph/0208512}}].

\bibitem{Ama2010}
R.~{Amanullah}, C.~{Lidman}, D.~{Rubin}, G.~{Aldering}, P.~{Astier},
  K.~{Barbary}, M.~S. {Burns}, A.~{Conley}, K.~S. {Dawson}, S.~E. {Deustua},
  M.~{Doi}, S.~{Fabbro}, L.~{Faccioli}, H.~K. {Fakhouri}, G.~{Folatelli}, A.~S.
  {Fruchter}, H.~{Furusawa}, G.~{Garavini}, G.~{Goldhaber}, A.~{Goobar}, D.~E.
  {Groom}, I.~{Hook}, D.~A. {Howell}, N.~{Kashikawa}, A.~G. {Kim}, R.~A.
  {Knop}, M.~{Kowalski}, E.~{Linder}, J.~{Meyers}, T.~{Morokuma}, S.~{Nobili},
  J.~{Nordin}, P.~E. {Nugent}, L.~{{\"O}stman}, R.~{Pain}, N.~{Panagia},
  S.~{Perlmutter}, J.~{Raux}, P.~{Ruiz-Lapuente}, A.~L. {Spadafora},
  M.~{Strovink}, N.~{Suzuki}, L.~{Wang}, W.~M. {Wood-Vasey}, N.~{Yasuda}, and
  T.~{Supernova Cosmology Project}, {\it {Spectra and Hubble Space Telescope
  Light Curves of Six Type Ia Supernovae at 0.511 $<$ z $< $1.12 and the Union2
  Compilation}},  {\em \apj} {\bf 716} (June, 2010) 712--738,
  [\href{http://arxiv.org/abs/1004.1711}{{\tt arXiv:1004.1711}}].

\bibitem{Gol2001}
M.~{Goliath}, R.~{Amanullah}, P.~{Astier}, A.~{Goobar}, and R.~{Pain}, {\it
  {Supernovae and the nature of the dark energy}},  {\em \aap} {\bf 380} (Dec.,
  2001) 6--18, [\href{http://arxiv.org/abs/astro-ph/0104009}{{\tt
  astro-ph/0104009}}].

\end{thebibliography}\endgroup

\end{document}